\input mn

\pageoffset{-2.5pc}{0pc}

\pagerange{1}
\pubyear{1996}
\volume{000}

\begintopmatter  
\title{History of the star formation in the local disk from the G dwarf 
metallicity distribution}
\author{H. J. Rocha-Pinto and W. J. Maciel}
\affiliation{Instituto Astron\^omico e Geof\'{\i}sico, Av. Miguel 
Stefano 4200, 04301-904 S\~ao Paulo SP, Brazil}
	       
\shortauthor{H. J. Rocha-Pinto and W. J. Maciel}
\shorttitle{The disk star formation history}

\acceptedline{Accepted . Received .}

\abstract {We address the question of the occurrence of star formation 
bursts in the solar neighbourhood by using the metallicity distribution 
of the G dwarfs. We present a method to recover the star formation 
history using simultaneously the metallicity distribution and the 
age--metallicity relation. The method associates the number of stars in 
a given metallicity interval with the corresponding time interval 
predicted by the age--metallicity relation. We take into account 
corrections relative to observational errors, cosmic scatter and scale 
height effects. The method is tested by simulations of the chemical 
evolution of a galaxy with irregular star formation rate, and our 
results show that it is possible to recover the star formation 
history fairly well. The application of the method to the solar 
neighbourhood shows evidences for at least two strong events of star 
formation: a burst 8 Gyr ago, and a lull 2-3 Gyr ago. These 
features confirm some previous studies on the star formation history 
in the local disk.}


\maketitle  %  finish the two spanning material

\section{Introduction}

The star formation rate (hereafter SFR) is one of the main parameters 
for galactic evolution studies. Previous attempts to recover the SFR 
history in the solar vicinity yield a present relative birthrate 
$b(T_G)$ close to unity, where $T_G$ is the present time, and $b(T_G)$ 
is the ratio between the present SFR, $B(T_G)$, and the average past 
SFR, $\langle B\rangle$. This fact has led many authors to conclude that
the SFR history in the solar neighbourhood was almost constant, or 
slightly decreasing over the lifetime of the Galaxy (see Scalo 1986, 
for a review). 

However, $b(T_G)\sim1$ indicates only that $B(T_G) = 
\langle B\rangle$, and not that $B(t)=\langle B\rangle$ at any time 
$t$, an expected result on the assumption that the SFR has been 
constant. There is no specific information on the past behaviour of 
the SFR in the function $b(T_G)$. Particularly, there is no indication 
that the SFR must be a smooth monotonic function of time, a hypothesis 
generally adopted for its computational simplicity or by the 
theoretical faith in the power of self-regulation in complex systems 
(Noh \& Scalo 1990).

In fact, several evidences strongly suggest that the local SFR has 
been quite irregular, with well-marked periods of enhancement and 
quiescence extending for about 1-3 Gyr. These evidences come from 
several independent studies relative to (i) stellar age distributions 
(Twarog 1980, revisited by Noh \& Scalo 1990; Meusinger 1991a); (ii) 
Ca II emission in late-type dwarfs (Barry 1988; Soderblom, Duncan \&
Johnson 1991); (iii) distribution of Li abundances (Scalo 1995, 
private communication); (iv) present-day mass function (Scalo 1987), 
(v) stellar kinematics (G\'omez et al. 1990; Marsakov, Shevelev \& 
Suchkov 1990), and (vi) white dwarf luminosity function (Noh \& Scalo 
1990; see also D\'{\i}az-Pinto et al. 1994). 

The major results emerging from these studies were summarized by 
Majewski (1993). He recognizes the occurrence of at least three star 
formation bursts, named as burst A, B and C. Burst A is a present star 
formation burst, while bursts B and C occurred approximately 5-6 Gyr 
and 8 Gyr ago, respectively. Between bursts A and B there is a long 
era of quiescence in the star formation activity, which is probably 
reflected in the Vaughan-Preston gap in the distribution of 
chromospheric emission among late-type dwarfs (Vaughan \& Preston 1980; 
Barry 1988; Henry et al. 1996).

The occurrence of bursts is likely to leave signatures in the 
metallicity distribution of G dwarfs, in the sense that there will be 
more stars with the mean metallicity corresponding to the time of the 
burst. In the present work, we make an effort to recover the SFR 
history from these possible signatures in the metallicity distribution 
of nearby long-lived stars, as recently derived by us (Rocha-Pinto \& 
Maciel 1996). In section~2, we present the method to recover the 
SFR history. Section~3 describes the simulations we have performed in 
order to test our method. In section~4, we present the results from 
the application of the method to the solar neighbourhood, and our main 
conclusions.

\section{A method to recover the SFR}

We want to relate the number of stars born in the time interval 
$(t,t+dt)$ with the total number of stars observed at a time $T$, 
which have metallicities between $(z,z+dz)$. The first function 
corresponds to the absolute SFR, $B(t)$; the second corresponds to 
the metallicity distribution of the region studied at time $T$, 
denoted as $N(z\vert T)$.

Let $Z(t)$ be the mean gas metallicity, which defines an 
age--metallicity relation (AMR) for the system under study. The number 
of stars born in the interval $(t,t+dt)$ with metallicities in 
$(z,z+dz)$ is
$$d^2n(z,t)=B(t)P(z,t)\,dt\,dz,\eqno\stepeq$$
where $P(z,t)$ is a probability function for a star born at $t$ to 
have a metallicity $z$. According to Tinsley (1975) and Basu \& Rana 
(1992),
$$P(z,t) = {1\over\sigma\sqrt{2\pi}} \exp\left[-{[z-Z(t)]^2\over 
2\sigma^2}\right], \eqno\stepeq$$
where $\sigma$ is the cosmic abundance scatter in the interstellar 
medium at birth. Integrating eq. (1) over time we have,
$$N(z|T)\,dz = {1\over\sigma\sqrt{2\pi}} \int^T_0 B(t) 
\exp\left[-{[z-Z(t)]^2\over 2\sigma^2}\right] \,dt\,dz.\eqno\stepeq$$
It can be seen from eq. (3) that $B(t)$ can be obtained if we have 
the metallicity distribution $N(z|T)$ and the age--metallicity 
relation $Z(t)$. 

Assuming that the AMR is a monotonous function of time, and that there 
is no cosmic scatter in the interstellar medium, each star born at $t$ 
would have a well defined metallicity $Z(t)$, so that we can write for 
a time interval $\Delta t$
$$B(t)\Delta t = N(Z)\Delta Z,\eqno\startsubeq$$
where $\Delta Z$ is the corresponding metallicity variation. Eq. (4a)
becomes for $\Delta t\to 0$,
$$B(t)\,dt = N(Z)\, dZ.\eqno\stepsubeq$$
This procedure is analogous to assign a {\it chemical age} to each 
star, defined as the age for which the mean gas metallicity is $Z$. 
The recovery of the SFR history in our method is based on a 
distribution of chemical ages.

However, {\it there is} a real cosmic scatter in the interstellar 
medium and, as a consequence, the metallicity $z$ of a star is 
generally different from the mean gas metallicity $Z$ at birth. In 
order to improve the distribution of chemical ages, we need to 
transform the observed metallicity distribution $N(z|T)$ in a 
distribution of mean metallicities at birth, $N(Z|T)$. Hereafter, for 
the sake of simplicity, we will omit the epoch $T$ relative to the 
metallicity distribution, which will be noted simply as $N(z)$. 
Substitution of eq. (4b) into eq.(3) yields
$$N(z)={1\over\sigma\sqrt{2\pi}}\int^{Z(T)}_{Z(0)} 
N(Z)\exp\left[-{(z-Z)^2\over 2\sigma^2}\right]dZ,\eqno\stepeq$$
where we have also omitted the time dependence of the metallicity. 
Eq. (5) shows that $N(z)$ is the convolution of the desired function 
$N(Z)$ with a cosmic scatter gaussian function.

Eq. (5) is much simpler than eq. (3). The method to solve it is 
the same used by Rocha-Pinto \& Maciel (1996) to correct the observed 
metallicity distribution by observational errors and cosmic scatter, 
and was first used by Pagel \& Patchett (1975). The observed 
metallicity distribution must be approximated by a gaussian function 
for deconvolution purposes. The mean metallicity distribution is 
obtained by 
$$N(Z)=\left.N(z)+\delta N_0(z)\right\vert_{z=Z},\eqno\stepeq$$
where
$$\delta N_0(z) = G^0_1(z)-G^0_0(z),\eqno\stepeq$$
and $G^0_0$ is the gaussian function adjusted to $N(z)$, and $G^0_1$ 
is its deconvolution, assuming $\sigma =0.20$~dex.   

That formalism assumes that the metallicity distribution is 
complete and representative of the chemical evolution of the system. 
To eliminate some possible bias from scale height effects, we must 
consider the Sommer-Larsen (1991) $f$-factors. In this case, the mean 
metallicity distribution is given by
$$N(Z)=\left.{N(z)\over f(z)}+\delta N_1(z)\right\vert_{z=Z},
\eqno\stepeq$$
where $f$ is the Sommer-Larsen factor, and
$$\delta N_1(z) = G^1_1(z) - G^1_0(z). \eqno\stepeq$$
$G^1_0$ is the gaussian function adjusted to $N(z)/f(z)$, and 
$G^1_1$ is its deconvolution, assuming $\sigma =0.20$~dex. Once we 
find $N(Z)$, the SFR history can be recovered by equation (4b).

\section{Simulating the SFR recovery}

We can test the validity of the assumptions of the method by means of 
simulations of the chemical evolution of a galaxy experiencing an 
irregular SFR. We assume a galaxy with no infall, and with a prescribed 
relative SFR, $b(t)$. The initial mass function $\xi(\log M)$ and 
$b(t)$ are connected by the procedures described by Miller \& Scalo 
(1979). We have used a present-day mass function similar to that of 
the solar neighbourhood (Scalo 1986), and the stellar lifetimes of 
Bahcall \& Piran (1983). 

\beginfigure*{1}
  \vskip 19cm
  \caption{{\bf Figure 1.} Simulations showing the original (solid 
  lines) and recovered (dashed lines) SFR.}
\endfigure

The functions $\xi(\log M)$ and $b(t)$ are suitably converted into 
their corresponding quantities according to the formalism developed
by  Tinsley (1980), $\phi(M)$ and $\psi(t)$, by the equations
$$\phi(M)={\log e\over {\cal M}}{{\xi(\log M)\over M}}\eqno\stepeq$$
and
$$\psi(t)={b(t)\over T_G}{\cal M},\eqno\stepeq$$
with
$${\cal M}=\log e\int^{{M_{max}}}_{{M_{min}}}\xi(\log M)\,dM,
\eqno\stepeq$$
where we take $M_{max}=62 M_\odot$ and $M_{min}=0.1 M_\odot$  
(Miller \& Scalo 1979). 

The equations for the evolution of the gas mass ($m_g$) and 
metallicity ($Z$) are given by
$${d\over dt}m_g(t)=-[1-R]\psi(t);\eqno\stepeq$$
$${d\over dt}\left[m_g(t)\,Z(t)\right]= -Z(t) [1-R] \psi(t) + 
y [1-R] \psi(t),\eqno\stepeq$$
where $y$ is the metal yield, and $R$ is the mass fraction returned to
the interstellar medium. 

The AMR for this galaxy is directly obtained by solving the above 
equation for $Z(t)$. The metallicity distribution is derived by the 
following procedure. Assuming a sample comprising $N_{\rm tot}$ stars, 
the number of stars born at $(t,t+\Delta t)$ is
$$B(t)\Delta t = b(t) {N_{\rm tot}\over T_G}\Delta t. \eqno\stepeq$$
The time interval $\Delta t$ corresponds to a metallicity interval 
$\Delta Z(t)$. For each star born in $(t,t+\Delta t)$, we associate 
a randomly chosen metallicity between $[Z(t), Z(t) + \Delta Z(t)]$. 
To simulate the intrinsic cosmic scatter, this metallicity is further 
shifted by a random gaussian deviation $\sigma=0.20$ dex.

We have used several input parametrizations for the SFR history 
corresponding to cases a--h, as shown in Fig. 1. Each parametrization 
is used in the derivation of the age--metallicity relation and the 
metallicity distribution. These last two functions are used in order 
to recover the original SFR history. The recovered SFR histories 
are also shown in Fig 1, compared with the original histories.

Fig. 1 includes several examples, ranging from a constant SFR (fig. 1a)
to several burst or series of bursts (figs. 1b-g), and also including
epochs of particularly low SFR (figs. 1g,h). We see from Fig. 1 that 
there is a good agreement between the original SFR histories (solid 
lines) and the recovered ones (dashed lines) in the whole time 
range, except at the extremes $t=0$ and $t=13$ Gyr. Fig. 1a 
corresponds to an originally constant SFR. The recovered SFR is 
nearly constant, showing a certain degree of noise. This could 
mask less intense star formation events, making them indistinguishable 
from a constant star formation activity. 

The bursts of Figs. 1b,c occur at the same time, but the intensity of 
the burst in Fig. 1c is six times lower than in Fig. 1b, and its 
duration is almost half the duration of the larger burst. While the 
recovered SFR in Fig. 1b does indicate the occurrence of the burst, 
in Fig. 1c it is barely noticeable. The method is likely 
to smear out the star formation events, in the sense 
that large amplitude bursts are recovered as broad-looking less 
intense features. As a consequence, we expect that the method has a 
finite resolving power, that is, star formation events occurring in a 
very short timescale cannot be recovered. This is illustrated by 
Figs. 1e and 1f. The original SFR in Fig. 1e is composed by two bursts 
with their maxima spaced by 2 Gyr, the second being less intense than 
the first. The recovered SFR combines these bursts together in a 
large star formation event. The second, less intense burst is noticeable
by a small bump at $t\sim 8$ Gyr (Fig. 1e, dashed line). Fig. 1f shows 
a similar case, where the  bursts are spaced by 3 Gyr. In this case, 
we can successfully isolate both bursts, suggesting that the resolution 
of our method is about 1.5 Gyr.

A feature shared by all recovered SFR histories deserves more 
attention. All histories show a burst at $t=13$ Gyr. This burst 
is probably an artifice created by the method, implicit in the counting 
procedure that  transforms the metallicity distribution in a 
distribution of chemical ages [equation (4b)]. The method assumes 
that all stars with $z > Z(T_G)$ were born at $T_G$, which largely 
overestimates the real number of stars born at $T_G$. 

\beginfigure{2}
  \vskip 9.5 cm
  \caption{{\bf Figure 2.} Age -- metallicity relations from the 
literature.}
\endfigure

A similar problem occurs in the recovery of $b(t)$ at $t=0$. All
SFR histories, except that in Fig. 1d, show a scarcity of stars born at 
$t=0$ relative to the original histories. In fact, the number of stars 
born at $t=0$ does not include the stars with ${\rm [Fe/H]}<-1.2$, 
which are removed from the sample, according to the chemical 
criterion adopted by Rocha-Pinto \& Maciel (1996), in order to 
separate halo and disk objects. As a conclusion, we expect the present 
method to produce correct results with a resolution of about 1.5 Gyr 
during the whole time span of galactic evolution, {\it except} at the 
extremes $t = 0$ and $t = 13 $ Gyr.

\section{Application to the local disk}

\subsection{The age-metallicity relation}

For the recovery of the SFR history in the local disk, we have used 
the metallicity distribution of G dwarfs recently derived by Rocha-Pinto 
\& Maciel (1996). Unfortunately, there is no well-established AMR in the 
literature. There are many determinations of this relation, but they are 
generally not consistent with each other, so that we decided to take
into account several age--metallicity relations. These include the
empirical AMR of Twarog (1980), Carlberg et al. (1985), Meusinger,
Reimann \& Stecklum (1991), Edvardsson et al. (1993), and  the 
parametrization of the AMR given by Rana (1991). All these relations 
are shown in Fig. 2, where we take the present time as 13 Gyr.
To allow their use in our method, all empirical AMRs were 
parameterized by an arbitrary increasing function of time.

The AMR by Carlberg et al. (1985) is in remarkable disagreement with 
the others concerning the early times of galactic evolution, suggesting 
that the gas was initially very rich. There is a generalized suspicion 
that this relation is biased toward old metal-rich stars (Meusinger et 
al. 1991; however, see Strobel 1991 and Marsakov et al. 1990). We 
decided to consider the SFR history derived from this relation as 
less conclusive.

\beginfigure*{3}
  \vskip 9.5 cm
  \caption{{\bf Figure 3.} Derived SFR histories in the local disk for 
  each adopted AMR. Also shown are the location of the peaks of
  bursts A, B and C (Majewski 1993), and the region corresponding to
  the Vaughan-Preston gap.}
\endfigure

The AMR by Edvardsson et al. (1993)  is a byproduct of a comprehensive
analysis of spectroscopic abundances of 189 F and G stars. The
curve shown in Fig. 2 corresponds to average points in 1 Gyr
bins, and generally agrees with the AMRs by Twarog (1980, 329 stars)
and Meusinger et al. (1991, 536 stars), except at times roughly in the
range 7--11 Gyr, when the AMR by Edvardsson et al. (1993) gives lower
abundances by approximately 0.1 dex.

From the work of Edvardsson et al. (1993), the cosmic scatter in the
insterstellar medium varies between 0.10 to 0.28 dex, with a most 
probable value of 0.20 dex. This value agrees well with a previous 
estimate by Twarog (1980). In our work, we will generally adopt a 
somewhat lower value, $\sigma=0.15$ dex. For a given AMR, the method for
the recovery of the SFR masks almost completely the chemical evolution
if we assume an abundance scatter of the same order of the 
dispersion in the metallicity distribution, which is $\sim$ 0.20 dex
according to Rocha-Pinto \& Maciel (1996).

\beginfigure{4}
  \vskip 9.5 cm
  \caption{{\bf Figure 4.} Comparison of the results using the AMR of
Edvardsson et al. (1993) and different values for the cosmic abundance 
scatter.}
\endfigure

\subsection{The local SFR history}

In Fig. 3, we show the SFR histories resulting from the use of 
the adopted AMR, again with the present time corresponding to 
$t = T_G = 13$ Gyr. We also show the location of the peaks of 
bursts B and C (Majewski 1993), and the region corresponding to the 
Vaughan-Preston gap.

Several remarks can be made on the recovered histories. First, we will
consider the histories derived from the AMR relations by Meusinger et
al. (1991), Twarog (1980), and Rana (1991), which show a better  
agreement with each other. In general, these histories show the
occurrence of an extended
enhanced star formation era 6 to 10 Gyr ago. We can also see a ``lull'' 
between 1 to 3 Gyr ago, when the star formation activity has decreased 
considerably. The results also indicate the occurrence of an intense burst 
at $t=T_G$. However, from the discussion in the previous section, we 
expect most of this burst to be an artifice produced by our method.
It should also be noted that the history from Rana's AMR cannot
be taken independently from that derived from Twarog's AMR, as the
parameterization presented by Rana (1991) is based on the data by Twarog
(1980).

The large star formation event present in these histories at 4 to 5 Gyr 
(8 to 9 Gyr ago) clearly corresponds to burst C (Majewski 1993). 
There is also an indication of  another star formation enhancement 
at 7 to 8 Gyr (5 to 6 Gyr ago) Gyr, corresponding to burst B (Majewski 
1993).  This feature is very similar to the result of our simulations shown 
in Fig. 1e, where the occurrence of a less intense burst close to the 
main burst is recovered as a small bump.

The history derived from the AMR by Carlberg et al. (1985) shows an 
essentially flat SFR, apart from strong events at $t = 0$ and
$t = 13$ Gyr. We should recall that the AMR of Carlberg et al. 
(1985) predicts a high initial metallicity for the disk. By analogy 
with the problem at $T_G$, the method also assumes that all stars with 
[Fe/H] $\le {\rm [Fe/H]}_{t=0}$ were born at $t=0$. As long as 
${\rm [Fe/H]}_{t=0}$ is very high in this AMR, it can be understood
why this SFR history has so many stars at $t=0$. In view of the fact
that the AMR of Carlberg et al. (1985) is probably biased, we decided 
to disregard the corresponding SFR history.

A more complex situation occurs with the AMR by Edvardsson et
al. (1993). From the results of Fig. 3, the history derived from this
AMR seems to indicate a nearly constant SFR over the disk lifetime,
except that some remnants of the star formation events shown in the
other histories are still visible at about 5 and 10 Gyr. A more detailed 
analysis of this SFR history is shown in Fig. 4, where we compare the 
results using two different values for the cosmic scatter, namely
$\sigma=0.15$ and $\sigma=0.20$. In this figure, it can be seen more
clearly that some star formation enhancements are present, which
approximately correspond to bursts B and C (Majewski 1993). 
Therefore, the results derived from the AMR by Edvardsson et al. (1993)
do not contradict the results derived from the remaining AMRs, namely
those by Twarog (1980), Meusinger et al. (1991), and Rana (1991),
although the derived SFR enhancements are considerably smaller.
 
\begintable{1}
  \caption{{\bf Table 1.} Mean SFR history for the local disk.}
\halign{ \hfill #  \hfill & \qquad\hfill #\hfill  \cr
$t$ (Gyr) & $b(t)$ \cr
0 to 1 & $1.14\pm 0.20$ \cr
1 to 2 & $0.79\pm 0.15$ \cr
2 to 3 & $1.09\pm 0.15$ \cr
3 to 4 & $1.26\pm 0.22$ \cr
4 to 5 & $1.64\pm 0.39$ \cr
5 to 6 & $1.42\pm 0.10$ \cr
6 to 7 & $1.14\pm 0.24$ \cr
7 to 8 & $0.95\pm 0.08$ \cr
8 to 9 & $0.78\pm 0.13$ \cr
9 to 10 & $0.61\pm 0.25$ \cr
10 to 11 & $0.54\pm 0.28$ \cr
11 to 12 & $0.39\pm 0.25$ \cr
(12 to 13 & $1.28\pm 0.22$) \cr}

\endtable

We obtained an average SFR history using the derived histories,
excluding that derived from the data of Carlberg et al., as shown in
Table 1 and Fig. 5. The formal errors are shown, and the last temporal
bin is given within parentheses, owing to the limitations of our method.

\beginfigure{5}
  \vskip 9.5 cm
  \caption{{\bf Figure 5.} Comparison of our mean SFR history and 
previous determinations in the literature.}
\endfigure

Fig. 5 also shows a comparison of our average SFR history  and some 
previous determinations in the literature: Twarog (1980) and 
Barry (1988), as presented by Noh \& Scalo (1990); Meusinger (1991a), 
after transforming to relative SFR; and Soderblom et al. (1991), after 
corrections by Rana \& Basu (1992). Note that the main events found 
in our SFR history are also present in the other determinations. 
Particularly, we note a large star formation era from 3 Gyr to  9 Gyr, 
and a considerably decrease in the star formation activity between 10 
and 12 Gyr. 

The second burst at 8 Gyr, suggested by Fig. 4, which corresponds to 
Majewski burst B, is not reproduced in the SFR mean history, as it is 
not present in the histories derived from Meusinger et al. (1991) and 
Twarog (1980). However, we do not discard the possibility of its 
occurrence, as long as all other histories in Fig. 5 seem support it. 
It is possible that this burst is not so intense as to be resolved by 
our method.

\subsection{Concluding Remarks}

Our results are in very good agreement with some previous 
determinations of the star formation history. However, it must be 
stressed that the mean history we have obtained, as well as the other 
histories presented in Fig. 5, reflect the history of the star 
formation amongst late-type dwarfs. We have not made any assumptions 
regarding possible variations of the initial mass function (IMF)
during the star formation bursts. In the case that the IMF 
favours massive stars during the bursts (Mezger 1987), we would expect 
that the global SFR history should be very different from that we have 
recovered. As long as there are few evidences concerning variations 
in the IMF in these environments (Scalo 1986, 1987; Larson 1992), we 
believe that the history we have recovered is representative of 
the global SFR activity. 

There are many independent evidences for the occurrence of star 
formation bursts in the solar neighbourhood. As long as the stars 
presently observed in our vicinity are likely to have born at 
different galactocentric distances, the bursts we find among solar 
neighbourhood stars could have occurred over a larger part of the 
disk (Meusinger 1991b).

Majewski (1993) gives a scenario for the evolution of our Galaxy with 
these bursts. The author suggests that the burst occurs during a 
certain time after which the global gas heating, due to supernova 
explosions, inhibits further star formation. During an interval of a 
few Gyr, the hot gas is efficiently mixed, so that the next stellar 
generation (characterized by the next burst) has a greater mean 
metallicity and a smaller abundance scatter. This scenario can 
satisfactorily explain the chemical properties of the stars which 
define bursts A, B and C (Majewski 1993). Majewski also presents 
some similarities between his stellar population scheme as discrete 
burst populations and the classification system proposed by Roman 
(1950, 1952). In this classification, the disk late-type dwarfs are
subdivided into a spectral sequence composed by the groups weak CN, 
weak-line and strong-line. The existence of these discrete stellar 
groups is more easily explained by the assumption that the building of 
the disk was not a continuous process. In fact, Majewski points out 
that the chemical and kinematical properties of the stars in the 
groups weak CN, weak-line and strong-line are in remarkable agreement 
with the properties of the stars which define bursts C, B and A, 
respectively.

A previous explanation for the occurrence of bursts in the disk was 
given by Scalo (1987). He remarks that the similarities between the 
ages of the bursts in the disk and those in the Magellanic Clouds may 
suggest that the bursts could be the result of past interactions 
between the Clouds and our Galaxy. In fact, it is well known that 
close encounters between galaxies could trigger star formation bursts 
(Combes 1987). This hypothesis also seems very attractive, because 
orbital calculations show that the Clouds are presently near the 
perigalacticon (Lin, Jones \& Klemola 1995), which could have triggered 
burst A.

An usual criticism with respect to the occurrence of star formation 
bursts in the local disk comes from the traditional view that our 
galaxy is a typical spiral galaxy, where the SFR has been constant or 
slightly decreasing over its lifetime. In fact, the Galaxy is probably 
not a typical object, and should not be considered so. The existence of 
nearby satellite galaxies (the Magellanic Clouds) and the warp in the 
disk at larger galactocentric distances indicate that our Galaxy is 
rather an interacting galaxy, subject to bursts and other phenomena 
that occur in such galaxies.

\section*{Acknowledgements}

We thank M\'arcio Catelan and John M. Scalo for strong encouragement 
and helpful suggestions. We also thank an anonymous referee for some
useful comments on the Edvardsson et al. (1993) AMR. This work was
partially supported by CNPq and FAPESP.

\end
\section*{References}

\beginrefs

\bibitem Bahcall J.N., Piran T., 1983, ApJ, 267, L77
\bibitem Barry D.C., 1988, ApJ, 334, 446
\bibitem Basu S., Rana N.C., 1992, Ap\&SS, 196, 1
\bibitem Carlberg R.G., Dawson P.C., Hsu T., Vandenberg D.A. 1985, 
    ApJ, 294, 674
\bibitem Combes F., 1987, in Thuan T.X, Montmerle T., and  Tran Thanh 
    Van, eds., Starbursts and Galaxy Evolution, Editions Fronti\`eres, 
    Gif sur Yvette, 313
\bibitem D\'{\i}az-Pinto A.,Garc\'{\i}a-Berro E.,Hernanz M.,Isern J., 
    Mochkovitch R., 1994, A\&A, 282, 86
\bibitem Edvardsson B., Anderson J., Gustafsson B., Lambert D.L., 
    Nissen P.E., Tomkin J., 1993, A\&A, 275, 101
\bibitem G\'omez A.E., Delhaye J., Grenier S., Jaschek C., Arenou F., 
    Jaschek M., 1990, A\&A, 236, 95
\bibitem Henry T.J., Soderblom D.R., Donahue R.A., Baliunas S.L., 1996, 
    AJ, 111, 439
\bibitem Larson R.B., 1992, in Tenorio-Tagle G., Prieto M., S\'anchez 
    F., eds., Star Formation in Stellar Systems, Cambridge Univ. Press, 
    Cambridge, 125
\bibitem Lin D.N.C., Jones B.F., Klemola A.R., 1995, ApJ, 439, 652
\bibitem Majewski S.R., 1993, ARA\&A, 31, 575
\bibitem Marsakov V.A., Shevelev Yu. G., Suchkov A.A., 1990, Ap\&SS, 
    172, 51
\bibitem Mezger,  1987, in Thuan T.X, Montmerle T., and  Tran Thanh 
    Van, eds., Starbursts and Galaxy Evolution, Editions Fronti\`eres, 
    Gif sur Yvette, 3
\bibitem Meusinger H., 1991a, Ap\&SS, 182, 19
\bibitem Meusinger H., 1991b, Astron. Nachr., 312, 231
\bibitem Meusinger H., Reimann H.-G., Stecklum B., 1991, A\&A, 245, 57
\bibitem Miller G.E., Scalo J.M., 1979, ApJS, 41, 513
\bibitem Noh H.-R., Scalo J., 1990, ApJ, 352, 605
\bibitem Pagel B.E.J., Patchett B.E., 1975, MNRAS, 172, 13
\bibitem Rana N.C., 1991, ARA\&A, 29, 129
\bibitem Rana N.C., Basu S., 1992, A\&A, 265, 499
\bibitem Rocha-Pinto H.J., Maciel W.J., 1996, MNRAS, 279, 447
\bibitem Roman N.G., 1950, ApJ, 112, 554
\bibitem Roman N.G., 1952, ApJ, 116, 122
\bibitem Scalo J.M., 1986, Fund. Cosm. Phys., 11, 1
\bibitem Scalo J.M., 1987, in Thuan T.X, Montmerle T., and  Tran Thanh 
    Van, eds., Starbursts and Galaxy Evolution, Editions Fronti\`eres, 
    Gif sur Yvette, 445
\bibitem Soderblom D.R., Duncan D.K., Johnson D.R.H., 1991, ApJ, 375, 
    722
\bibitem Sommer-Larsen J., 1991, MNRAS, 249, 368
\bibitem Strobel A., 1991, A\&A, 247, 35
\bibitem Tinsley B.M., 1975, ApJ, 197, 159
\bibitem Tinsley B.M., 1980, Fund. Cosmic. Phys., 5, 287
\bibitem Twarog B.A., 1980, ApJ, 242, 242
\bibitem Vaughan A.H., Preston G.W., 1980, PASP, 92, 385
\bibitem Yoss K.M., 1992, AJ, 104, 327
\endrefs
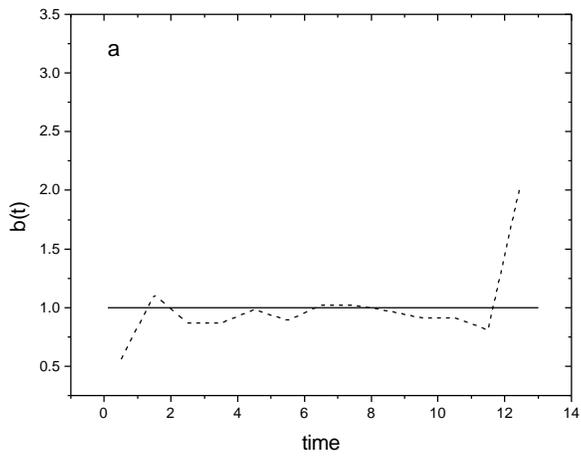
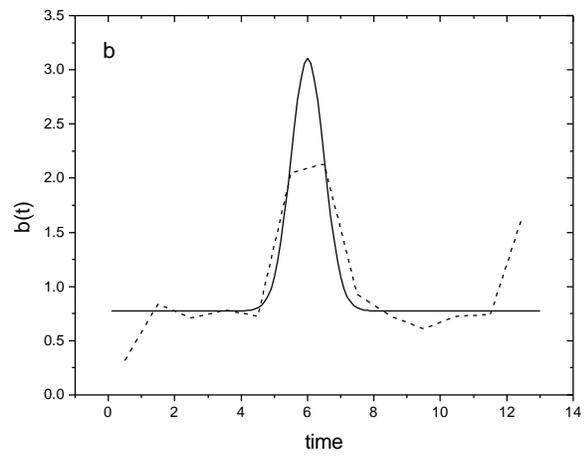
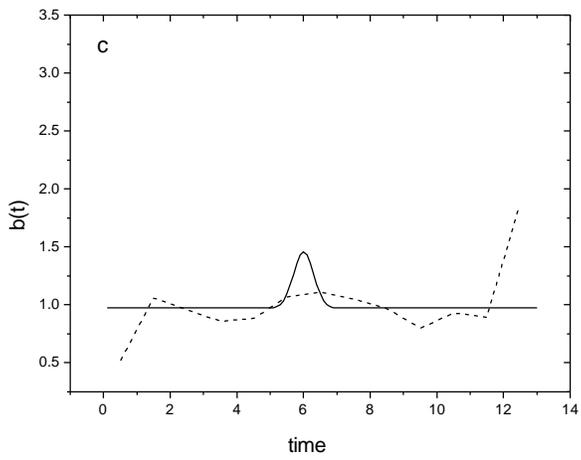
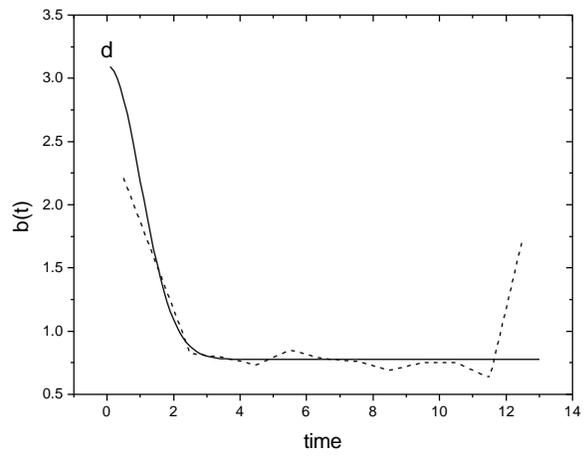
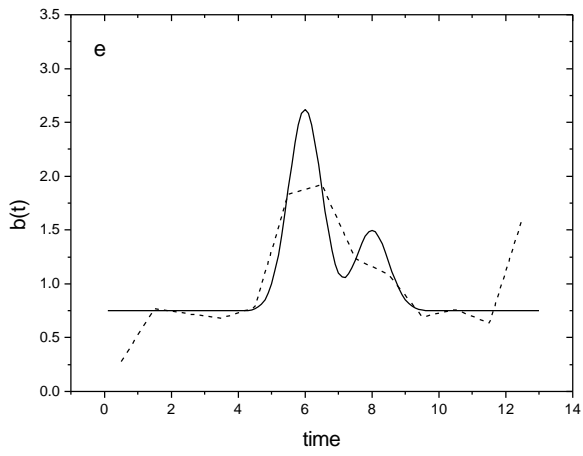
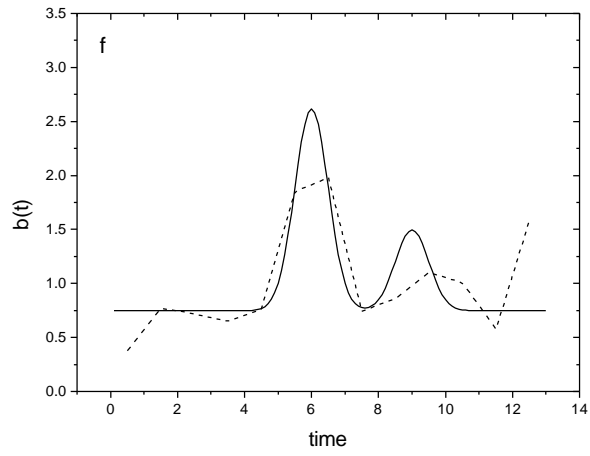
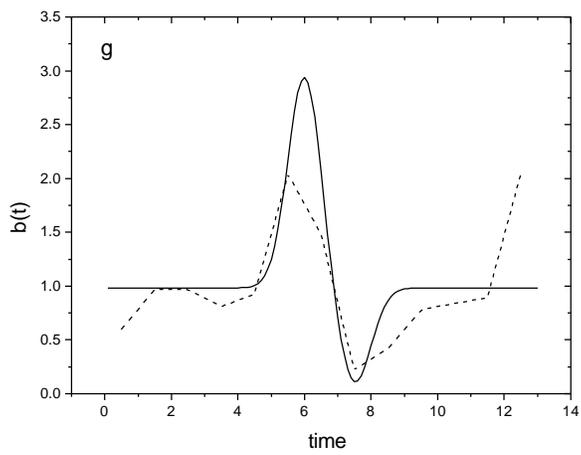
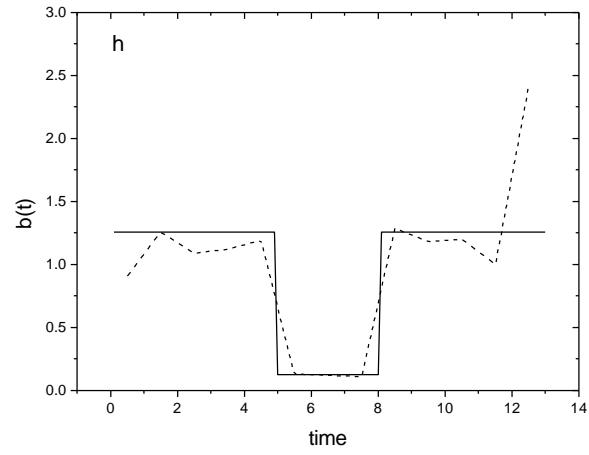
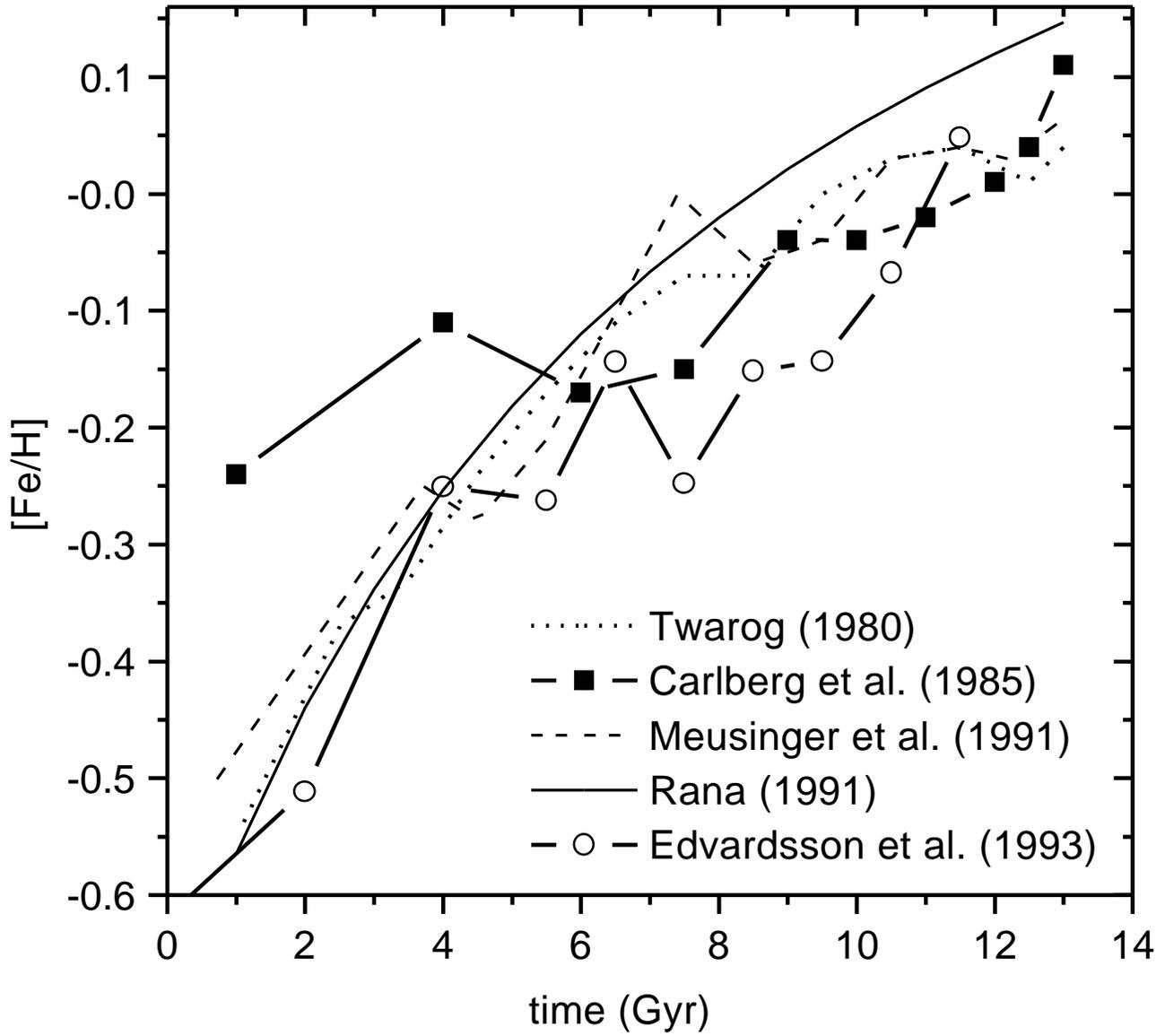
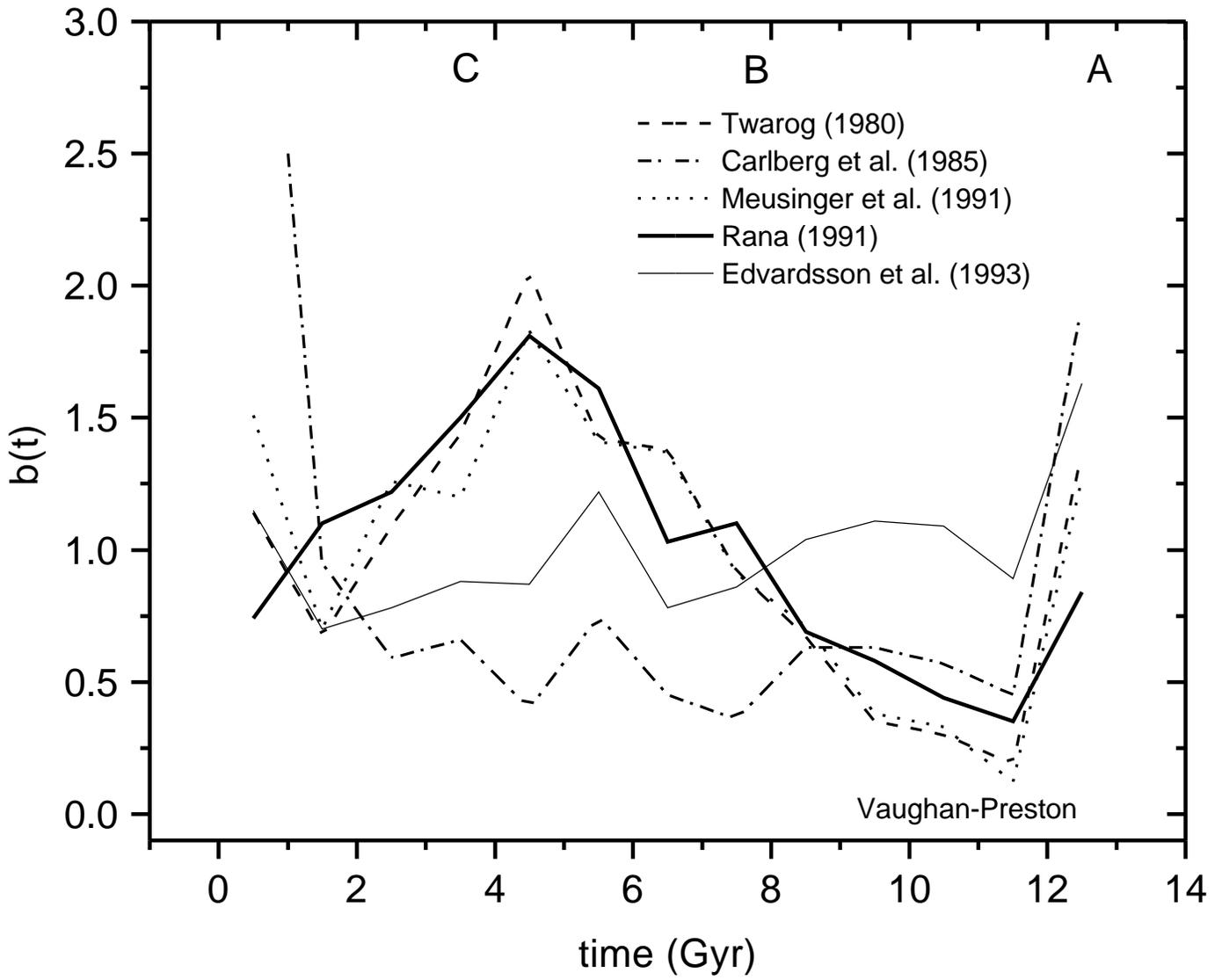
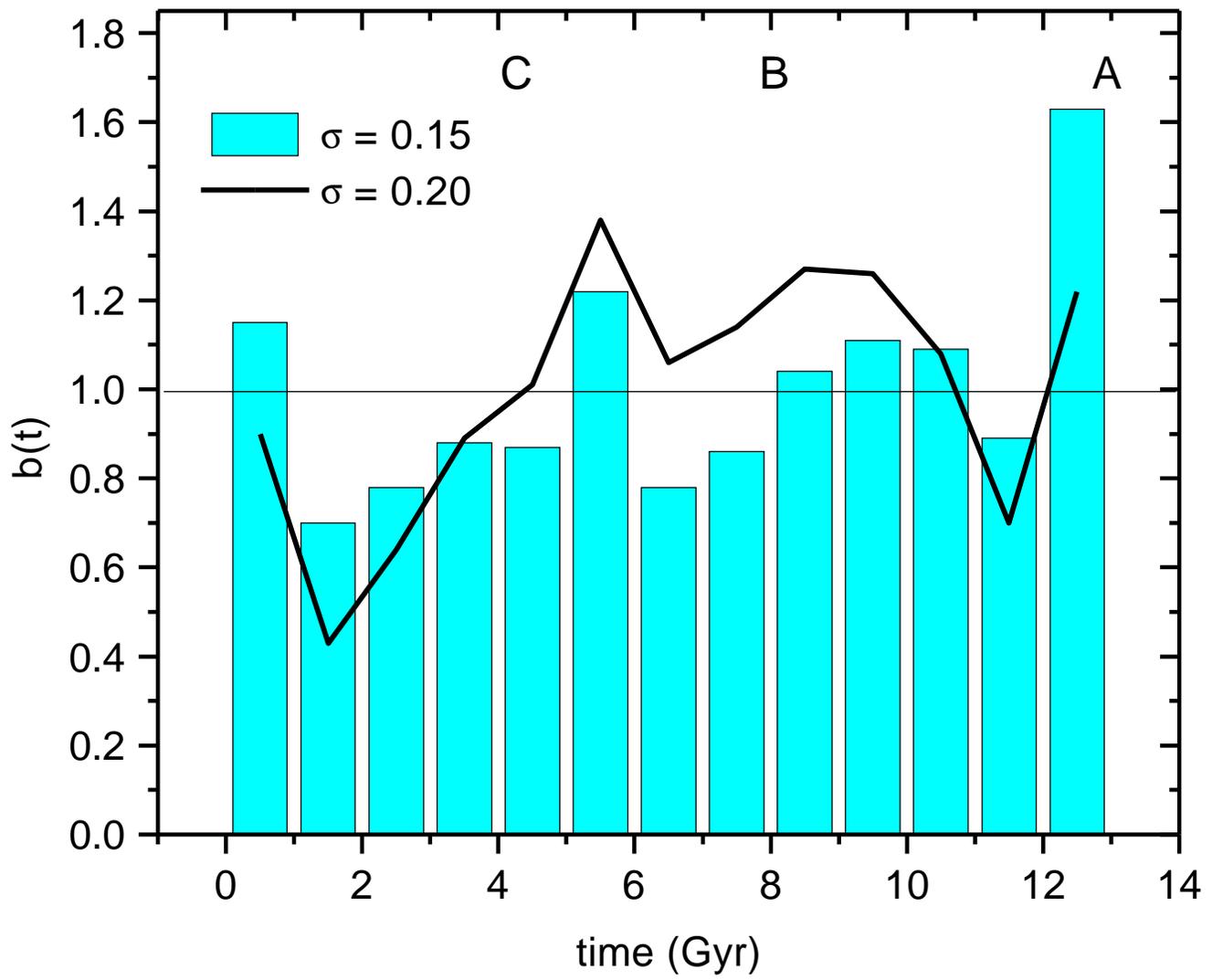
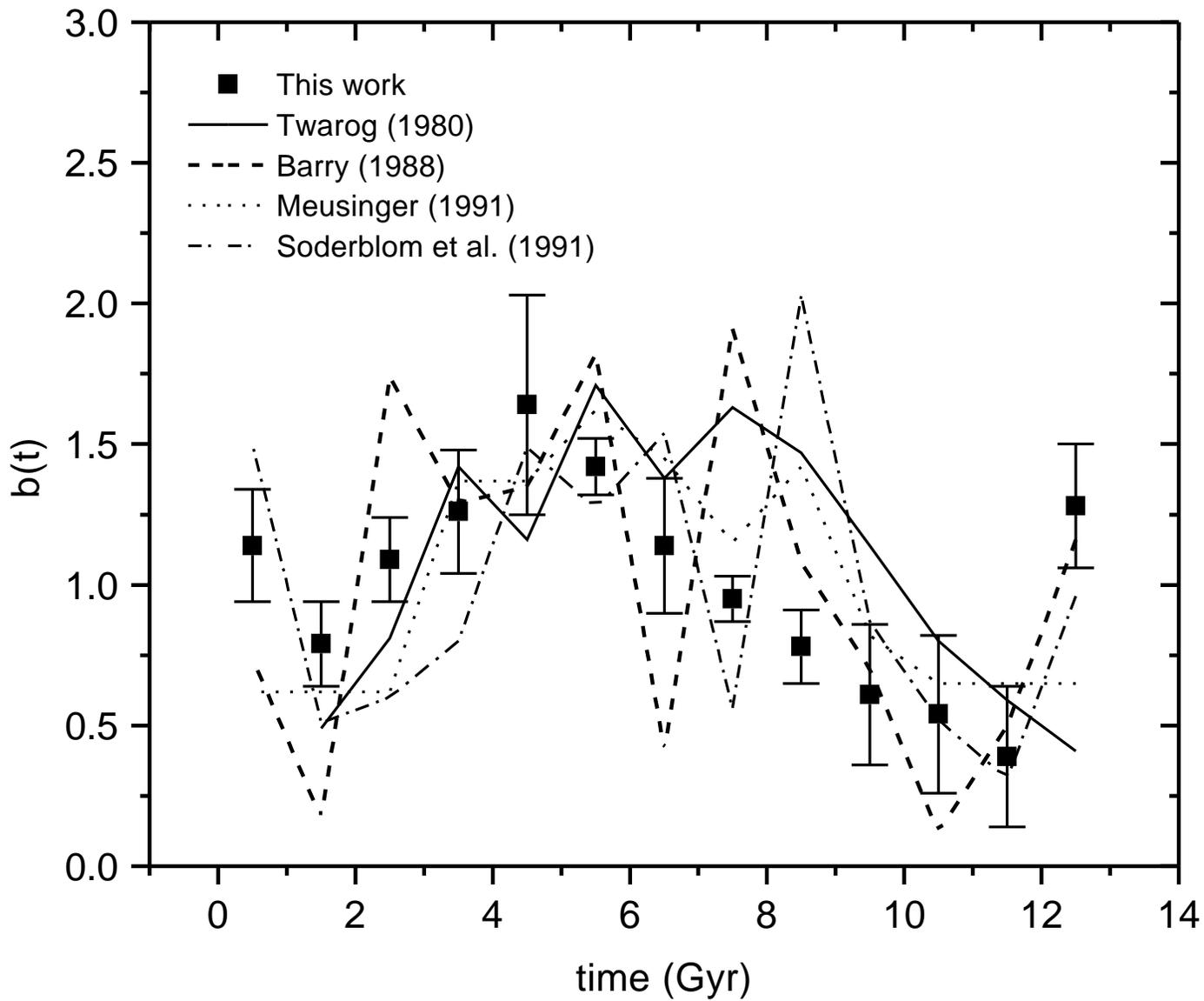
